\documentclass[12pt,preprint]{aastex}
\newcommand\etal{{\it et al.}}
\newcommand\lae{\mathrel{<\kern-1.0em\lower0.9ex\hbox{$\sim$}}}
\newcommand\gae{\mathrel{>\kern-1.0em\lower0.9ex\hbox{$\sim$}}}
\newcommand\kms{km~s$^{-1}$}
\newcommand\mone{$^{-1}$}

\newcommand\onezero{$1\rightarrow 0$}
\newcommand\twoone{$2\rightarrow 1$}
\newcommand\ammonia{NH$_3$}

\begin{document}

\title{A SEARCH FOR MOLECULAR GAS IN GHZ PEAKED SPECTRUM RADIO
SOURCES }

\author{Christopher P. O'Dea}
\affil{Department of Physics, Rochester Institute of Technology}
\authoraddr{  54 Lomb Memorial Drive, Rochester, NY 14623 \\
 email:  odea@cis.rit.edu }
\author{Jack Gallimore}
\affil{Bucknell University}
\authoraddr{Dept. of Physics, Lewisburg, PA 17837  \\
 email:  jgallimo@bucknell.edu }
\author{Carlo Stanghellini}
\affil{Istituto di Radioastronomia del CNR}
\authoraddr{ C.P. 141, I-96017 Noto SR, Italy \\ email:carlo@ira.noto.cnr.it }
\author{Stefi A. Baum }
\affil{Center for Imaging Science, Rochester Institute of Technology}
\authoraddr{  54 Lomb Memorial Drive, Rochester, NY 14623 \\
 email:	 baum@cis.rit.edu }
\author{James M. Jackson} \affil{Boston University}
\authoraddr{725 Commonwealth Ave, Boston, MA 02215 
\\ email: jackson@fish.bu.edu}

\begin {abstract}
We present searches for molecular gas (CO, OH, CS, and \ammonia) in six
GHz Peaked Spectrum (GPS) radio sources. We do not detect gas in any
source and place upper limits on the mass of molecular gas which
are generally in the range
$\sim 10^9$ to a few $\times 10^{10}$ M$_\odot$.  These limits are
consistent with the following interpretations: (1) GPS sources do not
require very dense gas in their hosts, \& (2) The GPS sources are
unlikely to be confined by dense gas and will evolve to become larger
radio sources.
\end{abstract}

\section{Introduction}

The GHz Peaked Spectrum (GPS) and Compact Steep Spectrum (CSS) radio
sources make up significant fractions of the extragalactic bright (cm
wavelength selected) radio source population ($\sim 10$\% and $\sim
20$\%) respectively, but are not well understood (e.g., O'Dea 1998).
They are powerful but compact radio sources whose spectra are
generally simple and convex with peaks near 1 GHz and 100 MHz
respectively. The GPS sources are entirely contained within the extent
of the nuclear narrow line region ($\lae 1$ kpc, NLR) while the CSS sources are
contained entirely within the host galaxy ($\lae 15$ kpc).

GPS and CSS sources are important because (1) they probe the NLR and
interstellar medium (ISM) of the host galaxy and (2) they may be 
the younger stages of
powerful large-scale radio sources -- giving us insight into radio
source genesis and evolution.

There are two main hypotheses for the GPS and CSS sources.  They could
be the young progenitor of the large scale powerful double sources 
(e.g., Carvalho 1985; Hodges \& Mutel 1987; Begelman 1996;
Fanti \etal\ 1995; Readhead \etal\ 1996b; O'Dea 1998; Snellen \etal\
2000; Alexander 2000). In this model they
propagate relatively quickly through the ISM of the parent galaxy with
advance speeds of a few percent of the speed of light. 
(Observed proper motions tend to
be a bit higher - in the range 0.1 - 0.2, e.g., Polatidis \& Conway 2003,
though the detections may be biased towards objects with the highest
velocities). In order to allow these high velocities, the ISM cannot be 
very dense and the total cold gas content can be no more than about 
$10^{10}$ M$_{\odot}$.  These sources must also undergo strong luminosity
evolution as they evolve, dimming by 1-2 orders of magnitude in radio
flux density.

In the second hypothesis, these sources are older and propagate much
more slowly though a dense ISM acquired via cannibalism (O'Dea, Baum,
\& Stanghellini 1991; De Young 1993; Carvalho 1994, 1998). These frustrated 
sources
interact strongly with their dense ambient medium driving a shock
which ionizes the gas and produces two effects -- (1) free free
absorption which is responsible for the turnover in the radio spectrum
and (2) optical emission lines (Bicknell, Dopita, and O'Dea 1997). In
this second model there is very little luminosity evolution. Bicknell
\etal\ predict that the host galaxies will be relatively gas-rich with
total masses of cold gas in the range $10^{10} - 10^{11}$
M$_{\odot}$. 
We note that this is similar to the cold gas content of
the Ultra Luminous Infrared Galaxies - ULIRG (e.g., Sanders, Scoville,
\& Soifer 1991). So far molecular gas (as traced by CO) has been detected 
in only one GPS source (1345+125 - Mirabel, Sanders, \& Kaz\`es 1989;
Evans \etal\ 1999b) with estimated mass of $6 \times 10^{10}$ M$_{\odot}$.
No  CO was detected in the GPS source 1934-638 at a limit of
$5\times 10^{10}$ M$_{\odot}$ (O'Dea \etal\ 1994b). 
 
Searches for the 21 cm line have produced a 50\% detection rate in
GPS and CSS sources while the detection rates for large sources
are only $\sim 10\%$ (Vermeulen \etal\ 2003; Pihlstr\"om, Conway
\& Vermeulen 2003; van Gorkom \etal\ 1989). 
This indicates that clouds of atomic hydrogen
are very common in the environments of GPS and CSS sources. 
Recent Hubble Space Telescope imaging and spectroscopy has 
shown high surface brightness emission line gas is aligned with the
radio axis in CSS radio galaxies (De Vries \etal\ 1997,1999, Axon
\etal\ 2000; O'Dea \etal\ 2002).  

Thus, the two models for GPS/CSS sources predict a substantial and testable
difference in the cold gas content of the host galaxies. And there is
some existing evidence that GPS/CSS sources contain dense gas in
their host galaxies. We have undertaken two complementary searches for 
molecular gas in GPS sources. First, we obtained
VLA observations to search for several molecular species (\ammonia, CS and
OH) in four GPS sources.
Second, we  obtained IRAM 30m observations of three
(relatively) low-redshift GPS sources to detect or set limits on the
cold gas content. These three sources are the lowest redshift sources
from volume limited subsets of complete samples of GPS (Stanghellini
\etal\ 1998) and CSS (Fanti \etal\ 1990) sources. One object, 0428+205 is
observed in both the VLA and IRAM searches. 

In this paper we present searches for molecular gas with the VLA and IRAM 
in 6 GPS sources.  
In principle, both emission and absorption searches are possible. 
Given the redshfits of
the sources and the cm wavelength flux densities, absorption and 
emission limits are
most sensitive for the VLA and IRAM data, respectively. 
We estimate constraints on the molecular gas content
and discuss the implications for models of these sources.

\section{Observations and Reduction}

\subsection{VLA Observations}

We searched for absorption in a transition of either ammonia, CS or OH
(whichever fell in a VLA band) in four GPS sources
(Table~\ref{tabsource}).  We observed with the VLA (Napier \etal\
1983) on January 22, 1993 in the A configuration in spectral line mode
(Mode 1A) using on-line hanning smoothing, 32 channels, channel
spacing of 390.625 kHz, and 12.5 MHz total bandwidth for each
observation. For the 15 GHz observations of 0237-233 we obtained
additional observations with the central frequency offset by $\pm 10$
MHz to cover a total bandwidth of about 33 MHz. Observational
parameters are given in Table~\ref{tabparam}.  Bandpass calibration
was performed using observations of the closest of either 3C84 or
3C454.3. Flux density calibration was carried out using observations
of 3C48. The data were reduced in AIPS following standard procedures.

A few very noisy channels at both ends of the spectra were deleted.
Since the sources are compact and unresolved at the VLA resolution we
used the task POSSM to average the data for all the antennas to obtain
the integrated spectrum. We subtracted a least squares linear fit to
the continuum.

\subsection{IRAM Observations}

We used the IRAM 30m millimeter-wave telescope, located on Pico
Veleta, Spain, to search for CO emission in the GPS sources 0116+319,
0428+205, and 0941$-$080. The observations took place over 11-13 July
1998 during daylight hours. Receivers were tuned to the redshifted CO
\onezero\ transition (115 GHz rest frequency) and, for 0116+319 only,
separate receivers were tuned simultaneously to the CO \twoone\
transition (230 GHz rest frequency). The beamsize at these transitions
is 21\arcsec\ and 11\arcsec, respectively. The telescope pointing and
focus were calibrated against scans of Jupiter, Mars, and the BL Lac
object 0235+164. To stabilize the spectral baselines and perform
initial sky subtraction, the targets were observed with a wobbling
secondary. The secondary throw angles ranged from 60\arcsec\ to
150\arcsec, and the wobble frequency was 0.25 Hz.

Each transition was observed in two backends, an autocorrelator with
1.25 MHz channel separation over a 600 MHz bandwidth, and a filterbank
with 1 MHz channel separation over a 512 MHz bandwidth. Spectral
baselines were subtracted using low-order polynomial fits to the raw
spectra. The baseline-subtracted spectra for each source were then
averaged using statistical ($1/{\rm RMS}^2$) weighting. The averaged
spectra were finally converted to mJy from $T_A$* using nominal
sensitivity curves provided by IRAM. The basic properties of the
reduced spectra are listed in Table~\ref{tabIRAMresults}.

\section{Results}

\subsection{VLA Results}

We did not detect any significant absorption in these sources. 
VLA Results are given in Table~\ref{tabresults}.  We obtain upper
limits to the absorption optical depth of typically a few percent with a total
range of 1-10\%.

The column density for OH$\lambda 1667$ absorption is given by
\begin{equation}
N(OH) \simeq 2.2\times 10^{14} T_{\rm ex} \tau \Delta V \ {\rm cm}^{-2}
\end{equation}
where $T_{\rm ex}$ is the excitation temperature, $\tau$ is the
optical depth, and $\Delta V$ is the FWHM of the line in km~s\mone\
(for 0428+205 and 2352+495 we adopt the values $\Delta V = 297$~\kms\
and $\Delta V = 82$~\kms, respectively, matching the widths of the
broadest HI absorption features found by Vermeulen \etal\ 2003).
Following O'Dea \& Baum (1987) we adopt a fiducial excitation
temperature $T_{\rm ex} = 10$~K (e.g., Dickey, Crovisier, \& Kaz\'es
1981; Turner 1985).  We convert from OH to H$_2$ column assuming a
relative abundance ratio $10^{-7}$ (e.g., Guelin 1985; Irvine \etal\
1985).  The mass of molecular gas is given by $M({\rm mol}) = 1.36 \pi
R^2 m_H2 N(H_2)$ where $m_H2$ is the mass of a hydrogen molecule, R is
the radius of the region considered (assuming for simplicity a plane
parallel geometry) and the factor of 1.36 includes the contribution of
He to the total molecular mass at solar abundance (e.g., Sanders
\etal\ 1991).

We applied a similar approach to evaluate the detection limits on the
CS and \ammonia\ absorption. The column density of CS can be estimated
by
\begin{equation}
N({\rm CS}) \simeq 3.6\times 10^{12} [1 - exp(-2.35/T_{\rm ex})] 
\tau \Delta V \ {\rm cm}^{-2}
\end{equation}
(e.g., Turner et al. 1973; Gardner \& Whiteoak 1978). The relative
abundance ratio CS / H$_2$ in dense Galactic cloud cores is $\sim
10^{-9}$ (e.g., Snell, Langer, \& Frerking 1982). As for OH, we also
adopt $T_{\rm ex} = 10$~K.

Only 0237$-$233 was searched for CS absorption (or emission). The
relatively narrow velocity coverage of the CS observations limits the
detection of absorption lines to line widths $\Delta V \la$~half of a
single bandpass. For the purposes of evaluating molecular mass limits,
we therefore adopted $\Delta V = 0.5\times$ the velocity range spanned
by an individual bandpass, or $\sim 120$~\kms. In principle, a large
amount of molecular gas could ``hide'' in a larger line-of-sight
velocity dispersion, but the range of velocities that absorption
selects is restricted by the compact and narrow background continuum
source. The velocity widths of absorption lines detected in active
spirals and ellipticals tend to be $\la 150$~\kms\ (e.g., Vermeulen et
al. 2003; Gallimore et al. 1999; van Gorkom et al. 1989; Dickey 1986;
Kaz\`es \& Dickey 1985), although very broad absorption lines have  
been detected (albeit rarely) in systems with larger background radio sources 
(e.g., PKS 2322-123, FWHM $\sim 735$ \kms\  - O'Dea, Baum, \& Gallimore 
1994; Taylor \etal\ 1999). It seems therefore unlikely that
large molecular columns are suppressed by line-of-sight velocity
dispersions (greatly exceeding 150~\kms) in this particular compact radio
source.

Assuming LTE, the column density of \ammonia\ is given by
\begin{equation}
N({\rm NH_3}) \simeq 2.8 \times 10^{13} T_{\rm ex} f_{11}^{-1}
\tau_{11} \Delta V \ {\rm cm}^{-2}
\end{equation} 
(e.g., Batrla, Walmsley, \& Wilson 1984; Herrnstein 2003) where
$\tau_{11}$ refers to the peak optical depth of the $(J,K)=(1,1)$
rotation inversion transition, and $f_{11}$ is the fraction of
\ammonia\ molecules in the $(1,1)$ state. Typical excitation
temperatures in Galactic cloud cores are $T_{ex} \simeq 40$~K (e.g.,
Morris et al. 1973; Barrett, Ho, \& Myers 1977; Ho et al. 1977), at
which $f_{11} \simeq 0.3$ (Herrnstein 2003). We adopt the relative
abundance ratio \ammonia\ / H$_2 = 10^{-8}$ (e.g., Morris et al. 1973;
Turner 1995). Only 0404+768 was searched for \ammonia\ absorption (or
emission), and for the purposes of evaluating the molecular mass
limit, we assume $\Delta V = 107$~\kms, based on the broadest HI
absorption line detected in this source (Vermeulen et al. 2003).

\subsection{IRAM Results}

The results of the IRAM observations are summarized in
Table~\ref{tabgas}.  We do not detect CO in any of the 3 GPS
sources searched. We obtain an upper limit to the integrated flux
density of the CO line $W_{\rm CO}$ and convert this to an upper limit
on the molecular gas content using the relation given by Sanders,
Scoville \& Soifer (1991)
\begin{equation}
M(H_2) = 1.18 \times 10^4 W_{\rm CO\ 1\rightarrow 0} D_{lum}^2 \ {\rm M}_\odot
\end{equation}
where $W_{\rm CO}$ is in units of Jy km s\mone\ and the luminosity
distance $D_{lum}$ is in units of Mpc. The conversion to molecular
hydrogen is subject to systematic uncertainties which have been
thoroughly discussed elsewhere (e.g., Bloemen \etal\ 1986; Scoville \&
Soloman 1987; Young \& Scoville 1991; Sanders \etal 1991; Maloney \&
Black 1988; Sage \& Isbell 1991). The limits derived from the CO
\twoone\ measurements assume CO \twoone\ / CO \onezero\ $ = 0.6$
(Lazareff et al. 1989). 

\section{IMPLICATIONS OF THE LACK OF DENSE MOLECULAR GAS}

We did not detect molecular gas in any of the 6 GPS sources we
searched.  The estimated upper limits to the masses of molecular gas
in the range few $\times 10^8$ to a few $\times 10^{10}$ M$_\odot$
(Table~\ref{tabgas}), with the weakest limit of $\la 10^{11}$ M$_\odot$
based on the non-detection of \ammonia\ absorption towards 0404+768.  
These are less than the amounts detected via CO in the most gas rich 
systems such as ULIRGs
(e.g., Sanders \etal\ 1991). However, most of these GPS sources could
still possess gas masses consistent with the lower end of the range
observed in luminous Infrared Galaxies. 

The detection of 21 cm H I absorption toward roughly half of the
known GPS sources, but the absence of molecular line absorption,
suggests two possibilities:  either (1) the circumnuclear gas
is predominantly atomic, or (2) atomic gas is easier to detect
than molecular gas.  For absorption experiments against a very
bright continuum source, only the line optical depth determines
the absorption signal strength.  Assuming LTE and large densities,
we can compare the column densities detectable by the two methods.

For atomic hydrogen 21 cm line, we can estimate the column density
by

$$N_H = 2 \times 10^{20} {\rm cm^{-2}} (\tau/0.1) (\Delta V/10 {rm km
s^{-1}}) (T/100 K)$$

where $N_H$ is the neutral atomic hydrogen column density, $\tau$ the
optical depth,
$\Delta V$ the linewidth, and $T$ the gas temperature.

For the CO 1-0 molecular line, the column density is
$$N_{H_2} = 5 \times 10^{21} {\rm cm^{-2}} (\tau/0.1) (\Delta V/10 {rm km
s^{-1}}) (T/100 K)$$

Here, we assume that the CO/H$_2$ abundance ratio is $10^{-5}$.  Results for
the
other molecules in our survey will give similar results.  Thus,
these estimates show that absorption atomic hydrogen is indeed easier to
detect than molecular line absorption.

The few powerful {\it extended\/} radio galaxies which have been detected
in molecular gas tend to have gas masses in the range $10^9 - 10^{10}$
M$_\odot$ (e.g., Israel \etal\ 1991; Mazzarella \etal\ 1993; O'Dea \etal\
1994a; Evans \etal\ 1999a; Lim \etal\ 2000; Das \etal\ 2004, in 
preparation). This suggests
that the GPS and larger radio galaxies may have similar molecular
gas content. 

These results suggest the following implications for GPS sources. 
\begin{itemize}
\item GPS sources do not require extremely dense environments.
\item The lack of very large masses of dense gas is consistent with the hypothesis
 that the majority of GPS sources
are generally not frustrated and will likely expand to become CSS sources. 
This is consistent with the proper motions of $\sim 0.1-0.2c$ observed in 
about ten GPS sources so far (e.g., Owsianik \& Conway 1998; Owsianik, 
Conway, \& Polatidis 1998; Tschager \etal\ 2000; see compilation by Polatidis \&
Conway 2003). However, these results do not {\it require\/} that all GPS sources
evolve to become large classical sources - though evolution is favored by the existing
data (e.g, O'Dea 1998). Alternate models are still possible for part of the 
GPS source population - e.g., some GPS sources may be intrinsically short lived 
(Readhead 1995).
\end{itemize}

\acknowledgements

We thank the referee for helpful comments. 
We are grateful to Eli Brinks for help with the OBSERVE file for the
VLA observations.  The VLA is operated by the US National Radio
Astronomy Observatory which is operated by Associated Universities,
Inc., under cooperative agreement with the National Science
Foundation. This work was supported by a grant from the STScI
Collaborative Visitor Fund.  IRAM is supported by INSU/CNRS (France),
MPG (Germany) and IGN (Spain).  This research made use of (1) the
NASA/IPAC Extragalactic Database (NED) which is operated by the Jet
Propulsion Laboratory, California Institute of Technology, under
contract with the National Aeronautics and Space Administration; and
(2) NASA's Astrophysics Data System Abstract Service.

\begin{deluxetable}{llrrllr}
\tablewidth{0pt}
\tablecaption{Source List  \label{tabsource}}
\tablehead{
\colhead{Name  } &
\colhead{Other Name }  &
\colhead{ID } &
\colhead{z } &
\colhead{D$_{\rm lum}$ } &
\colhead{Transition }&
\colhead{Rest Frequency}
\\
\colhead{ } &
\colhead{ } &
\colhead{ } &
\colhead{ } &
\colhead{Gpc } &
\colhead{ } &
\colhead{GHz }
\\
\colhead{ (1) } &
\colhead{ (2) } &
\colhead{ (3) } &
\colhead{ (4) } &
\colhead{ (5) } &
\colhead{ (6) } &
\colhead{ (7) }
}
\startdata
0116+319 & 4C31.04 & G &  0.0598  & 0.26  & CO 1-0   & 115.271  \\
.      &         &   &          &       &  CO 2-1   & 230.542  \\
0237-233 & OD263   & Q &    2.223   & 17.9 & CS 1-0   & 48.991   \\
0404+768 & 4C76.03 & G &    0.59846 & 3.51 & NH3(1,1) & 23.694  \\
0428+205 & OF247   & G &    0.219   & 1.07 & OH & 1.667   \\
.      &         &   &            &      & CO 1-0   & 115.271  \\
.      &         &   &            &      & CO 2-1   & 230.542  \\
0941-080 &         & G &   0.2280   & 1.12 & CO 1-0   & 115.271  \\
.      &         &   &            &      & CO 2-1   & 230.542  \\
2352+495 & OZ488   & G &    0.23831 & 1.18 & OH & 1.667   \\
\enddata
\tablecomments{(1) B1950 IAU Source Name. (2) Other Name. (3) ID, G=Galaxy,
i.e., narrow lines only, and Q=quasar, contains broad emission 
lines. (4) Redshift. 0404+768 and 2352+495 taken from Lawrence 
\etal\ (1996). 0428+205 from O'Dea \etal\ in preparation.  
(5) The luminosity distance estimated using Ned Wright's cosmology
calculator applet assuming
the current $\Lambda$ cosmology, i.e., H$_o = 71$ km s\mone\ Mpc\mone,
and $\Omega_M = 0.27$. 
(6) The molecular transition observed. (7) Rest frequency of transition. }
\end{deluxetable}

\begin{deluxetable}{lccccc}
\footnotesize
\rotate
\tablewidth{0pt}
\tablecaption{VLA Observation Parameters  \label{tabparam}}
\tablehead{
\colhead{Name  } &
\colhead{Integration Time }  &
\colhead{Central Frequency } &
\colhead{Velocity Resolution   } &
\colhead{Velocity Coverage  } &
\\
\colhead{ } &
\colhead{Min } &
\colhead{GHz } &
\colhead{\kms } &
\colhead{\kms }
}
\startdata
0237-233 &  39    & 15.2004  &   7.7  & 246 \\
0237-233 &  20    & 15.2104  &   7.7  & 246 \\
0237-233 &  24    & 15.2204  &   7.7  & 246 \\
0237-233 Total &  83    & 15.2104  &   7.7  & 650 \\
0404+768 &  43    & 14.82296 &    7.9 & 253 \\
0428+205 &  44    & 1.36781  &  85.5  & 2740  \\
2352+495 &  37    & 1.34790  &  86.7 & 2780  \\
\enddata
\tablecomments{(1) Source Name. (2) Integration time in minutes.
(3) Central frequency in GHz. (4) Velocity spacing of a 390 kHz
channel. (5) Total velocity coverage in \kms .
 }
\end{deluxetable}


\begin{deluxetable}{lllrrrc}
\tablewidth{0pt}
\tablecaption{IRAM  Observation Parameters  \label{tabIRAMresults}}
\tablehead{
\colhead{Source  } &
\colhead{Backend  }  &
\colhead{Transition  } &
\colhead{Int. Time  } &
\colhead{$T_A$* rms } &
\colhead{RMS     } &
\colhead{Channel width}
\\
\colhead{ } &
\colhead{ } &
\colhead{ } & 
\colhead{min  } &
\colhead{mK } &
\colhead{mJy } &
\colhead{km s$^{-1}$}
}
\startdata
 0116+319& 1 MHz  Filterbank &  CO \onezero& 309 &   2.0 &  12.6 & 2.9\nl
 .       & .                 &  CO \twoone & 309 &   5.5 &  52.8 & 1.5\nl
 .       & Autocorrelator    &  CO \onezero& 309 &   3.4 &  21.4 & 3.7 \nl
 .       & .                 &  CO \twoone & 309 &   4.0 &  57.6 & 1.8 \nl
 0428+205& 1 MHz  Filterbank &  CO \onezero& 720 &   1.6 &   9.6 & 3.9 \nl
 .       & Autocorrelator    &  CO \onezero& 720 &   1.5 &   9.0 & 4.8\nl
0941-080 & 1 MHz Filterbank  &  CO \onezero& 610 &   2.2 &  13.2 & 3.9\nl
 .       & Autocorrelator    &  CO \onezero& 610 &   2.0 &  12.0 & 4.9\nl
\enddata
\tablecomments{} 
\end{deluxetable}

\begin{deluxetable}{lccccc}
\footnotesize
\rotate
\tablewidth{0pt}
\tablecaption{VLA Results  \label{tabresults}}
\tablehead{
\colhead{Name  } &
\colhead{Flux Density }  &
\colhead{rms noise  } &
\colhead{optical depth } 
\\
\colhead{ } &
\colhead{Jy  } &
\colhead{mJy } &
\colhead{ } &
}
\startdata
0237-233 & 1.46   & 2.3  &  0.0047   \\  
0237-233 & 1.46   & 5.0  &  0.0103   \\
0237-233 & 1.46   & 3.4  &  0.0069   \\
404+768 & 1.45   &  2.7  &  0.0056   \\
0428+205 & 3.84   &  1.5  &  0.0012  \\ 
2352+495 & 2.71  &  2.2   &  0.0024  \\
\enddata
\tablecomments{(1) Source Name. (2) Flux density in Jy. 
(3) channel-to-channel rms noise in mJy. For 0237-233 the noise is given for
each of the three observations in order of increasing central frequency.   
(4) 3$\sigma$ upper limit on the 
optical depth of the line. 
 }
\end{deluxetable}

\begin{deluxetable}{lll}
\tablewidth{0pt}
\tablecaption{Constraints on Molecular Gas \label{tabgas}}
\tablehead{
\colhead{Name  } &
\colhead{Transition }&
\colhead{M(H$_2$) } 
\\
\colhead{ } &
\colhead{ } &
\colhead{ M$_\odot$ } 
\\
\colhead{ (1) } &
\colhead{ (2) } &
\colhead{ (3) } 
}
\startdata
0116+319 &  CO 1-0   & $5.9\times 10^8$   \\
0237-233 &  CS 1-0   & $5.9 \times 10^8$   \\
0404+768 &  NH3(1,1) & $1.2\times 10^{11}$   \\
0428+205 &  OH$\lambda 1667$  & $5\times 10^9$  \\
.      &  CO 1-0   &  $1.3\times 10^{10}$ \\
0941-080 &  CO 1-0   &  $2.0\times 10^{10}$  \\
2352+495 &  OH$\lambda 1667$ &  $2.8\times 10^9$ \\ 
\enddata
\tablecomments{(1) B1950 IAU Source Name. (2) The observed transition. (3) The
3$\sigma$ upper limit on the column density in that line. (4) The equivalent
upper limit to the molecular hydrogen column density. (5) The estimated upper
limit to the molecular gas mass. For the limits on absorption we assume
a size of $R=3$ kpc. 
}
\end{deluxetable}

\end{document}